\documentclass[%
 reprint,
superscriptaddress,
bibnotes,
 amsmath,amssymb,
prd
]{revtex4-1}

\usepackage{graphicx}
\usepackage{dcolumn}
\usepackage{bm}

\begin{document}


\title{Hyperbolic Self-Gravity Solver for Large Scale Hydrodynamical Simulations}

\author{Ryosuke Hirai}
\affiliation{Advanced Research Institute for Science and Engineering, Waseda University, 3-4-1, Okubo, Shinjuku, Tokyo 169-8555, Japan}
\author{Hiroki Nagakura}
\affiliation{Yukawa Institute for Theoretical Physics, Kyoto University, Oiwake-cho, Kitashirakawa, Sakyo-ku, Kyoto 606-8502, Japan}
\affiliation{TAPIR, Walter Burke Institute for Theoretical Physics, Mailcode 350-17, California Institute of Technology, Pasadena, CA 91125, USA}
\author{Hirotada Okawa}
\affiliation{Advanced Research Institute for Science and Engineering, Waseda University, 3-4-1, Okubo, Shinjuku, Tokyo 169-8555, Japan}
\affiliation{Yukawa Institute for Theoretical Physics, Kyoto University, Oiwake-cho, Kitashirakawa, Sakyo-ku, Kyoto 606-8502, Japan}
\author{Kotaro Fujisawa}
\affiliation{Advanced Research Institute for Science and Engineering, Waseda University, 3-4-1, Okubo, Shinjuku, Tokyo 169-8555, Japan}

\date{\today}

\begin{abstract}
A new computationally efficient method has been introduced to treat self-gravity in Eulerian hydrodynamical simulations. It is applied simply by modifying the Poisson equation into an inhomogeneous wave equation. This roughly corresponds to the weak field limit of the Einstein equations in general relativity, and as long as the gravitation propagation speed is taken to be larger than the hydrodynamical characteristic speed, the results agree with solutions for the Poisson equation. The solutions almost perfectly agree if the domain is taken large enough, or appropriate boundary conditions are given. Our new method can not only significantly reduce the computational time compared with existent methods, but is also fully compatible with massive parallel computation, nested grids and adaptive mesh refinement techniques, all of which can accelerate the progress in computational astrophysics and cosmology.
\end{abstract}

\maketitle

\section{Introduction}
Recent advances in theoretical astrophysics have mostly been led by the rapid progress in supercomputing, of the computers themselves and the numerical techniques. In particular, hydrodynamical simulations coupled with gravity have proved to be a powerful tool to reveal the dynamics of many astrophysical and cosmological phenomena such as supernovae, star formation, relativistic jets, accretion discs, formation of large scale structure etc \citep[]{kui10,nag13,tak14,hos15,tom15,ill15,hir15,fed16}. Most simulations deal with simplified models, assuming some symmetry and solving equations with reduced dimensions. In some fields, however, there are growing rationale that multidimensional effects can play a key role \citep[]{sai08,cou13b}, and some phenomena are essentially multidimensional \citep[]{sat15,ohl16}, meaning that numerical simulations also have to be carried out with full dimensionality. This itself can dramatically increase the numerical cost while at the same time, there are some studies where small scale effects can alter the global behaviour \citep[]{saw16}. In such cases it is necessary to resolve fine structures, making the calculation even more costly. Such computationally expensive calculations have become possible by making full use of state-of-the-art supercomputers, with the aid of a combination of efficient numerical schemes and parallelization technologies. However, computational resources for these large scale simulations are still limited, and it is often difficult to carry out systematic studies.

In astrophysical hydrodynamical simulations, it is usually not the hydrodynamics part that dominates the computational time. Instead, what prevents us from extending calculations to higher dimensions and higher resolutions, is the additional physics such as radiative transfer, nuclear reactions, neutrino transport, self-gravity etc. In order to carry out systematic studies in multi-dimensions, it is mandatory to construct rapid methods to treat these additional features. These additional effects are included by solving the governing equations of that feature and coupling it to the hydrodynamic Euler equations, or by applying approximated models based on feasible assumptions. For the case of self-gravity, the additional basic equation is the Poisson equation:
\begin{eqnarray}
 \Delta \phi=4\pi G\rho \label{Poissoneq.}
\end{eqnarray}
where $\Delta$ is the Laplace operator, $\phi$ the Newtonian gravitational potential, $G$ the gravitational constant and $\rho$ the mass density. This equation is an elliptic type partial differential equation (PDE) which can only be solved via direct matrix inversions or iterative methods or fast Fourier transform \citep[]{hes52,you54,fed62,sko75,bla75,rob82,kre83,van92,mul95,hua99,mat03,ric08,cou13a}. Despite the efforts made in the past few decades to construct rapid Poisson solvers \citep[]{hes52,you54,fed62,sko75,bla75,rob82,kre83,van92,mul95,hua99,mat03,ric08,cou13a}, it still remains the pain in the neck for many astrophysical hydrodynamic simulations. It becomes most problematic in multidimensional simulations with Eulerian schemes, and is sometimes approximated by monopoles even though the hydrodynamics are multidimensional \citep[]{han13,tak14,cou15,saw16}. The problem stems from the mathematical character of the equation itself, where the value on each cell depends on information from every other cell. This makes it extremely inefficient for parallelization, due to the huge amount of communication among memories and slowing down the whole calculation. The situation gets increasingly worse as the size of the simulation increases. 

On the other hand, the equation for general relativity is the Einstein field equations. When formulated as an initial value problem, the Einstein equations indicate that the evolution of gravitational fields are governed by a hyperbolic equation as long as it initially satisfies the Hamiltonian and momentum constraints \citep[]{mis73,alc08}. This implies that gravity is essentially hyperbolic, and its evolution only depends on its local neighbourhood. In this paper we propose a new method to circumvent the problems in Newtonian gravity, by incorporating the hyperbolicity of general relativity into the Poisson equation. Our new method can significantly reduce the computational cost of self-gravitational calculations. 

 Instead of the Poisson equation (\ref{Poissoneq.}), we choose to solve an inhomogeneous wave equation
\begin{eqnarray}
 \left(-\frac{1}{c_g^2}\frac{\partial^2}{\partial t^2}+\Delta\right)\phi=4\pi G\rho 
\label{HNeq.}
\end{eqnarray}
where we define $c_g$ as the propagation speed of gravitation. This equation was motivated from the essentially hyperbolic nature of gravity in general relativity. It roughly corresponds to the weak field limit of the Einstein equations. The Newtonian limit is achieved by assuming an infinite $c_g$, which is the cause of the difficulties, but here we just assume it is large, and not take the limit. Similar to electromagnetic fields, this equation will introduce causality, and the solution will therefore be somewhat like a retarded potential \citep[]{jac99}. In this way, Eq.(\ref{HNeq.}) can easily be parallelized since it is a hyperbolic PDE and only requires communication of memories between neighbouring cells. Our approach seems similar to the method introduced by \citet{bla75,kre83} where they convert the Poisson equation into a parabolic equation. However, the nature of a parabolic PDE and hyperbolic PDE is totally different, thus introducing different advantages and disadvantages to the method.

One important parameter that needs to be set is the value for the gravitation propagation speed $c_g$. A large enough $c_g$ will give us an equivalent solution to the Poisson equation, which is desired from the Newtonian point of view, but the computational time will be large due to the strict Courant-Friedrichs-Lewy condition. If we take a lower value for $c_g$, the computation will speed up, but the solution will deviate from that of the Poisson equation because the time derivative becomes comparable with the other terms. Thus the value for $c_g$ should be chosen carefully for each simulation according to the required accuracy and the computational resources available. Yet we show later in this paper that $c_g$ can be taken relatively small without affecting the solution, and can dramatically improve the numerical efficiency of self-gravity. This paper is organized as follows: in section 2, we will explain the numerical setups and methods used for our test calculations of our new method, the results will be shown in section 3 and we will discuss the errors in section 4. The conclusion will be given in section 5.

\section{Numerical Procedure}
We performed several calculations to demonstrate the efficiency of our new method. Firstly, we checked how well our new method maintains the equilibrium of a polytrope sphere in two-dimensions (2D) and three-dimensions (3D). Secondly, we simulate the head-on collision of equal mass polytropes in 2D. We use a hydrodynamical code which solves the ideal magnetohydrodynamic equations with the finite volume method, using the HLLD-type approximate Riemann solver \citep[]{miy05}. Since magnetic fields are ignored in our calculations, it is equivalent to using the HLLC scheme. Cylindrical coordinates are used for 2D  simulations assuming axisymmetry whereas 3D simulations are carried out in Cartesian coordinates. An ideal gas equation of state with an adiabatic index $\gamma=5/3$ is used for all calculations. An outgoing boundary condition is used for the outer boundaries.

 For self-gravity we solve two different equations; Eq.(\ref{HNeq.}) and the Poisson equation, for comparison. An iterative method called the MICCG method \citep[]{hes52,rob82} is used to solve the Poisson equation, with boundary values given by multipole expansion. Eq.(\ref{HNeq.}) is solved by simple discretization with the aid of the cartoon mesh method \citep[]{alc01} to simplify the cylindrical geometry in the 2D tests. Robin boundary conditions are applied for the outer boundary \citep[]{gus98}. As for the value of $c_g$, we normalize it by the characteristic velocity 
\begin{equation}
 c_g=k_g(c_s+|\bm{v}|)
\end{equation}
where $c_s$ is the sound speed, $\bm{v}$ is velocity, and $k_g$ is an arbitrary parameter that should be larger than unity. The timestep condition for the wave equation will become $k_g$ times stricter than for the hydrodynamical part. Although the gravity and hydrodynamical equations should essentially be solved simultaneously, here we choose to solve them on separate timelines. In this way, the wave equation will be solved $k_g$ times during one hydrodynamical timestep, and will reduce the computational cost. Owing to the fact that the wave equation only depends on the density distribution, and since the density distribution does not significantly change during one timestep, this will give sufficient accuracy. It should also be noted that the Courant number used to decide the timesteps for the gravity and hydrodynamical parts do not necessarily need to coincide. If we take larger Courant numbers for the gravity part than the hydrodynamics, the computational cost can be reduced even more. In this paper we simply take both Courant numbers to be 0.3, but the results did not change even for larger Courant numbers such as 0.9.

For the first test calculation, we place a polytrope sphere with a polytropic index $N=3$ at the centre of the 2D cylindrical grid. The sphere has a mass and radius of $(M, R)=(8 M_\odot, 3.75 R_\odot)$. The computational domain is taken approximately twice the stellar radius in both radial and longitudinal directions, and divided into  $(N_r\times N_z)=(210\times280)$ cells. A dilute atmosphere is placed around the star, with a mass negligible compared to the stellar mass. We simply wait for several dynamical times to see whether the star stays in mechanical equilibrium. Two simulations are carried out for comparison, one by solving the Poisson equation throughout (P model), and one by solving Eq.(\ref{HNeq.}) with $k_g=5$ (H model). The initial condition is given by solving the Poisson equation in both cases.

As a demonstration of 3D capabilities, we place the same polytrope sphere at the origin of a three-dimensional Cartesian grid. Plane symmetry is assumed for all three directions, which will leave us with an eighth of the star. The computational domain is taken $\sim1.5$ times the stellar radius in each direction, and divided into $(N_x\times N_y\times N_z=140^3)$ cells. The resolution of the grid is equivalent to the first test calculation. To make it a 3D specific problem, we add random density perturbations with an amplitude of $<1\%$. This will induce some stellar oscillation modes but overall, the star should stay in a stable state. Since we use a relatively large number of cells, it is extremely difficult to solve the Poisson equation. In fact, it was impossible on our workstation to solve in a realistic timescale, so we interpolate from the exact solution as an initial condition for the gravitational potential instead.

To test a more dynamically evolving case, we place another identical polytrope sphere $4.2\times10^{11}$ cm away from the centre of the region in the longitudinal direction on a 2D cylindrical grid. We assume equatorial symmetry, which mirrors the star on the opposite side. Since we do not give any orbital motions, the two stars will simply fall into each other by the gravitational force of each other, causing a head-on collision. Like in the first test, we carry out the simulation with the two different types of self-gravity for comparison, and call them the P and H models.

\section{Results}
\begin{figure}[tbp]
\includegraphics{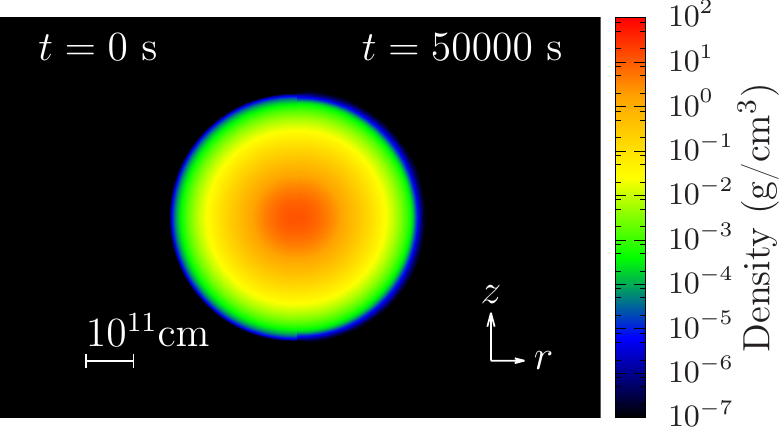}
\caption{\label{fig:statstar} Density plot for the stationary star test. Left panel: initial condition, Right panel: $5\times10^4$ s later. }
\end{figure}

\begin{figure}[tbp]
 \includegraphics{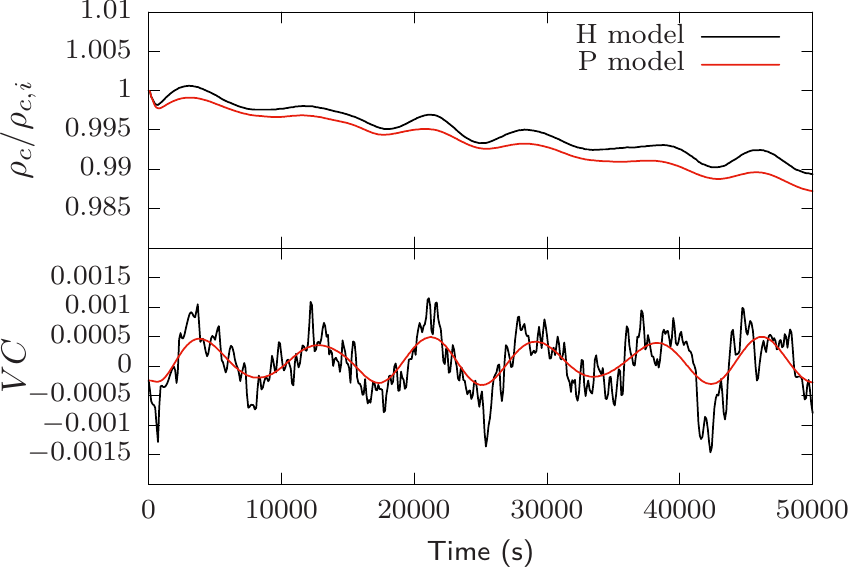}
 \caption{Evolution of the central density (Upper panel) and degree of satisfaction of Virial theorem ($VC$, lower panel) in the 2D static star simulations. Density is normalized by the initial central density, and $VC$ is defined in Eq.\ref{eq:VC}. Red lines: H model, Black lines: P model.\label{fig:rhocvirial}}
\end{figure}

Fig.\ref{fig:statstar} shows the density distribution of the initial condition on the left side, and $\sim5$ dynamical times later on the right side for the H model. Both panels show almost identical distributions, indicating that hydrostatic equilibrium of the star is well resolved with this grid. The degree of equilibrium can be checked in Fig.\ref{fig:rhocvirial}, which show the evolution of central density and the degree of satisfaction of the Virial theorem ($VC$) defined as 
\begin{eqnarray}
 VC=\frac{W+3(\gamma-1)U}{|W|}\label{eq:VC}
\end{eqnarray}
where $U$ and $W$ are the internal and gravitational energies integrated over the bound zones (zones with negative total energy). The initial condition for the polytrope sphere is given simply by interpolation from the exact solution. So as soon as the simulation starts, the star tries to adjust to its equilibrium condition on the discretized grid. This leads to a slow decrease in the central density, but the decline rate is extremely slow and it is safe to assume that the star is resolved properly on this grid, with both methods. There is a roughly dynamical timescale oscillation in the value of $VC$ in both P and H models. But the amplitude is very small and does not grow, which indicates that the star satisfies the virial equilibrium condition throughout the simulation.  The computational time was $\sim5$ times shorter for the H model than the P model.

\begin{figure}[tbp]
 \includegraphics{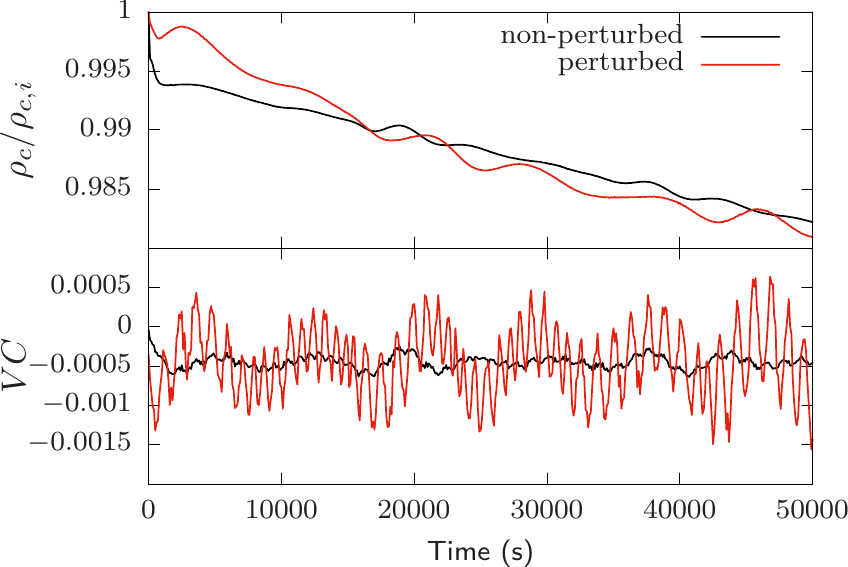}
 \caption{Evolution of the central density (Upper panel) and degree of satisfaction of Virial theorem ($VC$, lower panel) in the 3D simulations. Density is normalized by the initial central density, and $VC$ is defined in Eq.\ref{eq:VC}. Red lines: perturbed model, Black lines: non-perturbed model.\label{fig:3dvirial}}
\end{figure}

Similar results were obtained for the 3D star case, depicted in Fig.\ref{fig:3dvirial}. The black lines show the non-perturbed star case, which is simply an extention of the H model calculation to 3D and in different coordinates. It is remarkable that the star remains in virial equilibrium even in 3D, at a degree of $\sim0.05\%$. The red lines show the evolution of the same star with $\sim1\%$ random density perturbations. There is no notable difference in the evolution of the central density, only declining $\sim2\%$ after $\sim5$ dynamical times. The fluctuation around virial equilibrium is larger than the non-perturbed model, but does not grow in time, staying in a stably oscillating state at the same timescale as the 2D test.

\begin{figure}[tbp]
\includegraphics{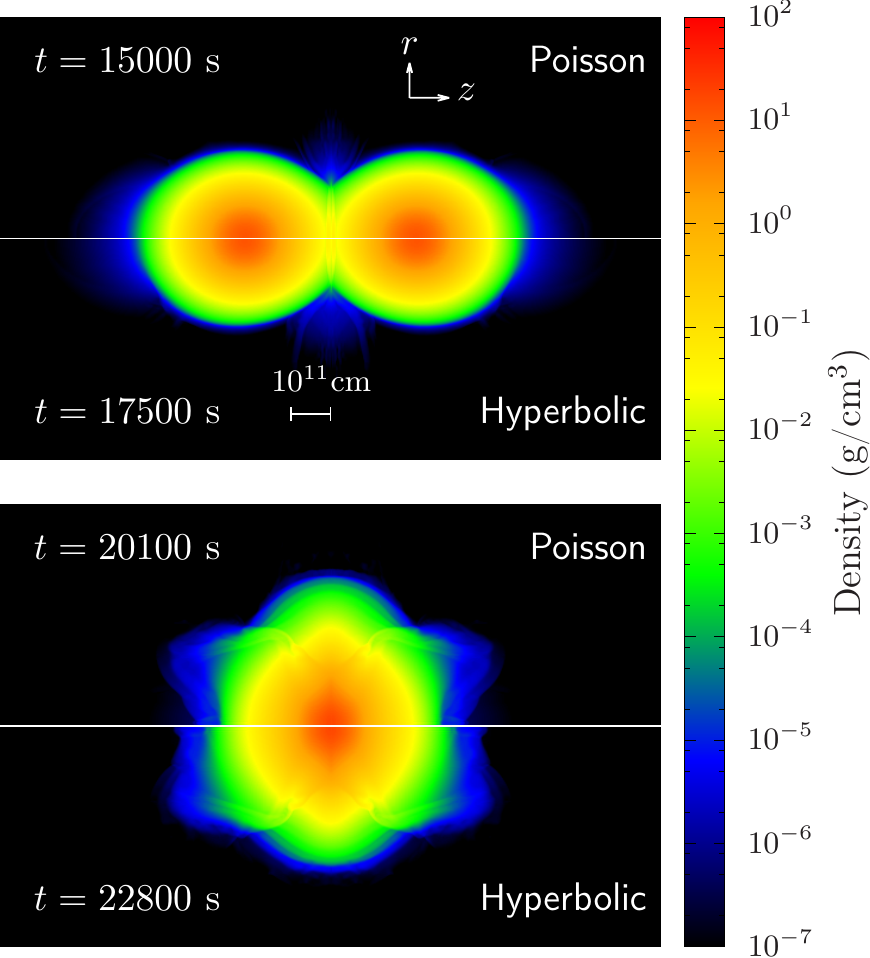}
\caption{\label{fig:headonsnap} Snapshots of the density distribution for the head-on collision simulations. Upper halves of each panel; P model, lower halves; H model. The time elapsed are written on the left corners, and white line at the centre shows the coordinate axis.}
\end{figure}

Fig.\ref{fig:headonsnap} shows the density distribution of the head-on collision simulations with the two different methods at two different times. The upper halves of each panel are results for the P model, and the lower halves are for the H model. It can be seen that the two stars fall into each other, causing a head-on collision, forming a shock at the interface. The stars then merge to become a single star, but a part of the envelope is blown away by the shock. Although the evolution is delayed by $\sim15\%$ in the H model, the overall behaviour of the dynamics between the two models are quite similar. This already indicates that our new method can at least be used for qualitative studies.  Moreover, the total computational time of the H model was $\sim30$ times shorter than the P model, implying that our new method is most efficient for dynamically evolving gravitational fields. This is because when the gravitational potential is moving, the MICCG method needs more iterations than stationary situations to converge to its solution.

\section{Discussions}

Here we are not interested in the physics of the test calculations carried out, but in the difference between the two methods. Our aim was to produce an efficient method to treat self-gravity, that reproduces the same results as with previous methods which solve the Poisson equation. In this section we quantitatively evaluate the differences between the solution obtained in our simulations by the new method and the solution of the Poisson equation. We focus on the head-on collision simulation, since it had the largest difference and because we are more interested in applying our new method to dynamically evolving systems.

One of the main causes of the difference between the two methods is the boundary condition. For the Poisson solver, we use Dirichlet boundary conditions with values given by multipole expansion, which obtains the exact solution for the Poisson equation for the given density distribution. On the other hand, the boundary condition used in the new method is a Robin boundary, which is equivalent to assuming monopole gravity. Since the higher order terms are non-negligible in the current situation, this boundary condition is inappropriate.

We carry out several additional simulations to quantify the effects of the boundary condition, and seek how to improve the results. The parameters used in our extra simulations are listed in Table \ref{tab:Models}. Eq.(\ref{HNeq.}) is used for self-gravity in all models. Model aR05 corresponds to the H model explained above. Three different modifications are made to single out the effects of the boundary condition. In the first approach, we apply the Dirichlet boundary condition by multipole expansion as in the P model (aD05). This will directly remove the boundary error, although the calculation becomes heavy and is inappropriate for practical use. Our second approach is to widen the computational domain without changing the resolution (aR05-cR05), which will weaken the multipole effects at the boundary. Finally, we also change the value of $k_g$ to a larger value (aR20-cR20), which should bring our equation closer to the Poisson equation. 

We define the relative ``error'' as 
\begin{eqnarray}
 \delta\tilde{\phi}(\bm{r}):=\frac{\phi_\textsc{h}(\bm{r})-\phi_\textsc{p}(\bm{r})}{\phi_\textsc{p}(\bm{r})}
\end{eqnarray}
$\phi_\textsc{h}$ is the gravitational potential calculated with our new method, and $\phi_\textsc{p}$ is the solution for the Poisson equation at the given density distribution. Hence the average relative error is
\begin{equation}
\left<\delta\tilde{\phi}\right>=\left(\frac{\int_\textsc{V} (\delta\tilde{\phi})^2dV}{\int_\textsc{V}dV}\right)^\frac{1}{2}
\label{eq:error}
\end{equation}
where the integrals are taken over the entire region. 

\begin{table}[tbp]
  \caption{Model Descriptions}
 \begin{ruledtabular}
  \begin{tabular}{ccccccc}
   Model& $r_\textrm{max}$\footnotemark[1](cm) & $z_\textrm{max}$\footnotemark[1](cm) & $N_r$\footnotemark[2] & $N_z$\footnotemark[2]& b.c.\footnotemark[3] & $k_g$ \\
   \tableline
   aD05  & $6.3\times10^{11}$ & $8.4\times10^{11}$ &210&280&Dirichlet&5 \\
   aR05  & $6.3\times10^{11}$ & $8.4\times10^{11}$ &210&280&Robin&5 \\
   bR05  & $1.8\times10^{12}$ & $2.4\times10^{12}$ &600&800&Robin&5\\
   cR05  & $3.6\times10^{12}$ & $4.8\times10^{12}$ &1200&1600&Robin&5\\
   aR20  & $6.3\times10^{11}$ & $8.4\times10^{11}$ &210&280&Robin&20 \\
   bR20  & $1.8\times10^{12}$ & $2.4\times10^{12}$ &600&800&Robin&20\\
   cR20  & $3.6\times10^{12}$ & $4.8\times10^{12}$ &1200&1600&Robin&20\\
  \end{tabular}\label{tab:Models}

 \end{ruledtabular}
 \footnotetext[1]{Size of the computational domain in each direction.}
 \footnotetext[2]{Number of zones in each direction.}
 \footnotetext[3]{Boundary conditions. Dirichlet boundaries are applied by multipole expansion.}
\end{table}

\begin{figure}[tbp]
\includegraphics{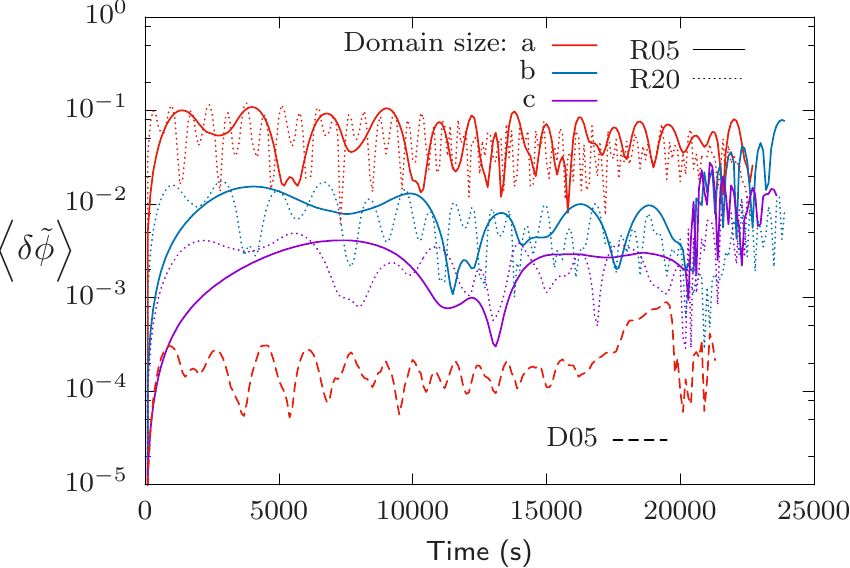}
\caption{\label{fig:diffevo} Time evolution of the average relative error for each model. Colours of the lines denote the domain size as described in Table \ref{tab:Models}. Dashed line: aD05 model. Solid lines: $k_g=5$ models, dotted lines: $k_g=20$ models.}
\end{figure}

Fig.\ref{fig:diffevo} shows the time evolution of the average relative error in each model. All lines fluctuate around a certain value, indicating that the error does not pile up in most cases. The maximum error was $\sim10\%$ even in our ``worst'' model (aR05, aR20). This is the cause of the $\sim15\%$ delay in the collision time. The error was reduced most when the Dirichlet boundary condition was applied (aD05; red dashed line), where the error does not exceed $\sim0.1\%$ throughout the calculation, and the delay time also became negligible. This is a surprisingly good agreement, and proves that the difference of our method to previous ones only arise from the boundaries. Our hypothesis is further verified by the other simulations with larger computational regions. The relative error is roughly inverse proportional to the number of zones, from $\sim10\%$ in $5.88\times10^4$ zones to $\sim0.003\%$ in $1.92\times10^6$ zones. The wider the region, the smaller the errors. This is because the relative contribution of the boundary to the computational domain is smaller for wider regions, and also the multipole effects are weakened at the boundary. Another fact to be noted is that the error does not strongly depend on the value of $k_g$ used in the simulation. The average error simply fluctuates around a value determined only by the domain size, at a frequency proportional to $\sim c_g/L$ where $L$ is the size of the domain. In fact, even if we take $k_g=2$, the overall behaviour is indistinguishable with other models as long as we take a large enough region. At the most turbulent and messy situations like after the collision ($t\gtrsim20000$ s), the errors rise higher in the lower $k_g$ models because they cannot react fast enough to rapidly evolving systems.

The reason for the oscillations in the errors can be understood by decomposing the gravitational potential into two parts $\phi=\phi_\textsc{p}+\phi_\textsc{e}$. Here we assume that $\phi_\textsc{p}$ is the solution for the Poisson equation ($\Delta\phi_\textsc{p}=4\pi G\rho$), and $\phi_\textsc{e}$ is the deviation from it. If we plug this in to Eq.(\ref{HNeq.}) and use the Poisson equation, we are left with
\begin{eqnarray}
 \left(-\frac{1}{c_g^2}\frac{\partial^2}{\partial t^2}+\Delta\right)\phi_\textsc{e}=\frac{1}{c_g^2}\frac{\partial^2}{\partial t^2}\phi_\textsc{p}\label{Erreq.}
\end{eqnarray}
This is the equation which describes the creation and propagation of the error $\phi_\textsc{e}$. If the initial condition satisfies the Poisson equation, i.e. $\phi_\textsc{e}(t=0)=0$, the only errors are generated by the source term on the right hand side and the boundary conditions. Besides the boundary, the source of the error is apparently the second time derivative of the gravitational potential, which is determined by the motion of the density distribution. This is why the error rised at the later times in Fig.\ref{fig:diffevo} where it was turbulent and messy. Due to the fact that this is a wave equation, any errors that are generated will propagate away out of the boundaries. The creation and propagation of errors is what causes the small oscillations of the errors in all of our test calculations. The amplitude of the errors are determined by the magnitude of this source term, which can be estimated by combining the Poisson equation, continuity equation and equation of motion. By taking the time derivative of the Poisson equation and using the continuity equation, one can get
\begin{eqnarray}
 \frac{\partial}{\partial t}\Delta\phi_\textsc{p}=-4\pi G\nabla\cdot(\rho \bm{v})
\end{eqnarray}
and then
\begin{eqnarray}
 \frac{\partial}{\partial t}\nabla\phi_\textsc{p}=-4\pi G\rho\bm{v}
\end{eqnarray}
Similarly by taking the time derivative again and using the equation of motion,
one can obtain something like
\begin{eqnarray}
 \frac{\partial^2}{\partial t^2}\nabla\phi_\textsc{p}=4\pi G((\rho\bm{v}\cdot\nabla)\bm{v}+\nabla p-\rho\nabla\phi_\textsc{p})
\end{eqnarray}
depending on the physics included. From this equation, it can easily be estimated that
\begin{eqnarray}
 \frac{\partial^2}{\partial t^2}\phi_\textsc{p}\sim\mathcal{O}(G\rho (c_s^2+\bm{v}^2))
\end{eqnarray}
So if we normalize Eq.(\ref{Erreq.}) by the original Eq.(\ref{HNeq.}), we can say that the relative amplitude of the error is roughly
\begin{eqnarray}
 \left|\frac{\phi_\textsc{e}}{\phi}\right|\sim\mathcal{O}\left(\frac{c_s^2+\bm{v}^2}{c_g^2}\right)
\end{eqnarray}
Provided that $c_g$ is taken larger than the characteristic velocity, or when there is not so much accelerating motion, the right hand side on Eq.(\ref{Erreq.}) can be assumed to be sufficiently small.

From the above results, we conclude that our new method can be safely used even for dynamically evolving systems provided that $c_g$ is chosen large enough and the outer boundary condition is given appropriately. Robin boundaries seem to be appropriate for any kind of application due to the fact that gravitational forces can be well approximated by monopoles at large distances from the source. It is also numerically efficient since it only requires information of the neighbouring cell. The only problem is that the boundary should be taken far enough from the source to reduce the errors sufficiently. Larger regions lead to larger computational cost, weakening the advantage of the new method. One possible workaround is to take a wider region only for the gravitational potential, and solving Eq.(\ref{HNeq.}) on an extended grid exceeding the region for hydrodynamics. If the density near the hydro boundaries are low enough, or in other words, the total mass inside the region is conserved, it will be possible to approximate that the extended regions are close to vacuum, and calculate the wave equation without the source term. The cost for the wave equation is cheap, thus we can extend the region relatively easily without increasing the total cost. Periodic boundaries are also suitable for this method whenever appropriate. In such cases, the average density of the computational region should be subtracted from the source term of Eq.(\ref{HNeq.}).

\begin{figure}[tbp]
\includegraphics{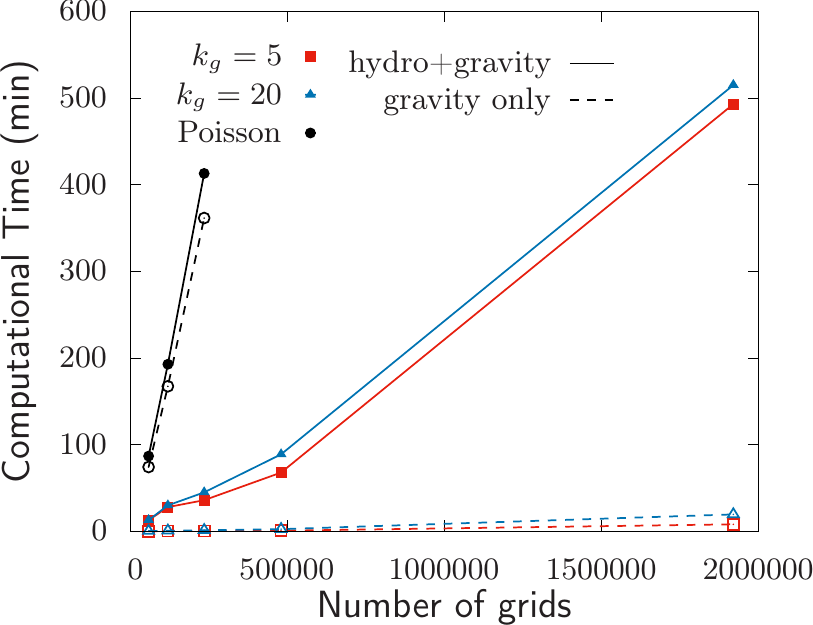}
\caption{\label{fig:comptime} Computational time until the stars contact for different methods. Square plots: R05 models, Triangle plots: R20 models, Circle plots: same calculation with Poisson solver. Solid lines: total computational time, Dashed lines: time for the gravity part only.}
\end{figure}

Fig.\ref{fig:comptime} shows the computational time it took until the stars come in contact ($\sim15000$ s) for each method. Calculations were carried out on a 172.8 Gflops machine with OpenMP parallelization on 8 threads. It can be seen that the computational time for the gravity part (dashed lines) can be dramatically reduced compared to previous methods, and the benefit becomes more prominent as the scale of the calculation increases. Since our test simulation was dominated by the gravity part with previous methods, the new method improved the overall performance directly. Almost $90\%$ of the computational time was spent on the gravity part using the Poisson solver, whereas the fraction is $\sim1\%$ with the new method. This is a remarkable improvement, since it is not so common with existing solvers that the time spent on the gravity solver is negligible compared to the hydrodynamics. For other cases where the computational time is dominated by other implementations, the improvement in the gravity part may not be so critical, e.g. in core collapse simulations which implement detailed microphysics, the fraction of time used for computing gravity is typically below $\sim10\%$, so the reduction of the total time will be at most $\sim10\%$. The computational time for the gravity part with this method scales linearly to the number of cells, which is much better than previous methods which usually scale as $\mathcal{O}(N^2)$ or $\mathcal{O}(N\log N)$. Multigrid methods are supposed to scale as $\mathcal{O}(N)$ too, but the absolute number of operations are obviously much smaller with the new method and much more simpler. Our method will suit even more on even larger scale simulations parallelized by MPI. In these cases the communication between memories are sometimes the bottleneck, but our new method will not be restricted by this since it does not require intensive communication. It should also be compatible for adaptive mesh refinement or nested grid techniques, and in this way, the outer boundary can be taken far enough without significantly increasing the computational cost.

\section{Conclusion}
A new method has been introduced to treat self-gravity in Eulerian hydrodynamical simulations, by modifying the Poisson equation into an inhomogeneous wave equation. As long as the gravitation propagation speed is taken to be larger than the hydrodynamical characteristic speed, the results agree with solutions for the Poisson equation depending on the boundary condition. If the errors from the boundary are removed in some manner, by applying Dirichlet boundaries or placing the boundary far away, the solution almost perfectly satisfy the Poisson equation. The computational time of the gravity part was reduced by an order of magnitude, and it should become more prominent for larger scale simulations. It is also fully compatible for numerical techniques such as parallelization, nested grids, adaptive mesh refinement, extending its superiority over existent methods.

The sole parameter that needs to be set is $c_g$, the gravitational propagation speed. This should ideally be taken as the speed of light, but our test simulations suggest that it can be taken to fairly small values as long as it exceeds the characteristic velocity of the hydrodynamics. Considering the computational time it is good that we can take it fairly small, but the effects on the errors should be clarified in future studies.

\begin{acknowledgments}
This work was supported by the Grants-in-Aid for the Scientific Research from the Ministry of Education, Culture, Sports, Science, and Technology (MEXT) of Japan (NoS. 24103006, 24740165, and 24244036), the HPCI Strategic Program of MEXT, MEXT Grant-in-Aid for Scientific Research on Innovative Areas ``New Developments in Astrophysics Through Multi-Messenger Observations of Gravitational Wave Sources'' (Grant Number A05 24103006), and the Research Grant for Young Scientists, Early Bird Program from Waseda Research Institute for Science and Engineering. H.N. was supported in part by JSPS Postdoctoral Fellowships for Research Abroad No. 27-348.
\end{acknowledgments}

\end{document}